\documentclass[prl,twocolumn,groupedaddress]{revtex4}
\usepackage{graphicx,color}
\usepackage{amssymb}   % for math
\usepackage{amsmath}
\usepackage{epstopdf}
\usepackage[bookmarks=false]{hyperref}
\usepackage{bm}

\begin{document}

\title{Generating Giant Spin Currents Using Nodal Topological Superconductors}
\author{Noah F. Q. Yuan, Yao Lu, James J. He and K. T. Law}
\thanks{phlaw@ust.hk}

\affiliation{Department of Physics, Hong Kong University of Science and Technology, Clear Water Bay, Hong Kong, China }

\begin{abstract}
In this work, we show that a giant spin current can be injected into a nodal topological superconductor, using a normal paramagnetic lead, through a large number of zero energy Majorana fermions at the superconductor edge. The giant spin current is caused by the selective equal spin Andreev reflections (SESAR) induced by Majorana fermions. In each SESAR event, a pair of electrons with certain spin polarization are injected into the nodal topological superconductor, even though the pairing in the bulk of the nodal superconductor is spin-singlet s-wave. We further explain the origin of the spin current by showing that the pairing correlation at the edge of a nodal topological superconductor is predominantly equal spin-triplet at zero energy. The experimental consequences of SESAR in nodal topological superconductors are discussed.
\end{abstract} 

\maketitle

%\section{Introduction}
{\bf Introduction}--- The search for Majorana fermions in condensed matter systems has been an important topic in recent years \citep{Wilczek,Kane,Qi,Alicea_rev,Beenakker_rev}. This search is strongly motivated by the fact that Majorana fermions are non-Abelian particles and have potential applications in quantum computation \citep{Kitaev,Alicea_braiding,Ivanov}. Recently, it was further pointed out that Majorana fermions, due to their self-Hermitian properties, could induce spin currents in paramagnetic leads \citep{SESAR,Wu, XinLiu} and correlated spin currents in spatially separated leads \citep{James_BDI, Oreg}. These properties make it possible for Majorana fermions to have potential applications in superconducting spintronics \citep{Eschrig, Linder}.

In particular, it was pointed out that a single Majorana end state of a topological superconducting wire can induce the so-called selective equal spin Andreev reflection (SESAR) at the normal lead/topological superconductor (N/TS) interface \citep{SESAR}. In SESAR processes, only electrons with certain spin polarization $\bm {n}$ in the normal lead can couple to the Majorana fermion and undergo Andreev reflections. Importantly, the reflected holes are due to missing electrons with the same spin polarization $\bm {n}$ below the Fermi energy. As a result, two electrons with equal spin tunnel into the superconducting wire and form a spin-triplet Cooper pair in each Andreev reflection event. On the other hand, electrons with opposite spin polarization $-\bm {n}$ in the normal lead are decoupled from the Majorana fermion and get reflected as electrons with unchanged spin. Therefore, a spin current with spin polarization in the $\bm {n}$ direction can be injected into the superconductor using a \emph{paramagnetic} lead. At the same time, a spin current is generated in the lead.

In this work, instead of studying isolated Majorana modes, we study the spin transport properties of the 2D nodal topological superconductor with a large number of spatially overlapping zero energy Majorana modes at the sample edge \citep{Tanaka_Nodal,Sato_MFB,Schnyder_Nodal,Chris_MFB}. The zero energy edge modes are associated with Majorana flat band (MFB), which connects the nodal points of the superconductor in the projected band structure. These flat bands are analogous to the surface Fermi arcs connecting the Weyl points in Weyl semimetals \citep{WSM1,WSM2}. 

\begin{figure}
\centering
\includegraphics[width=3.2in]{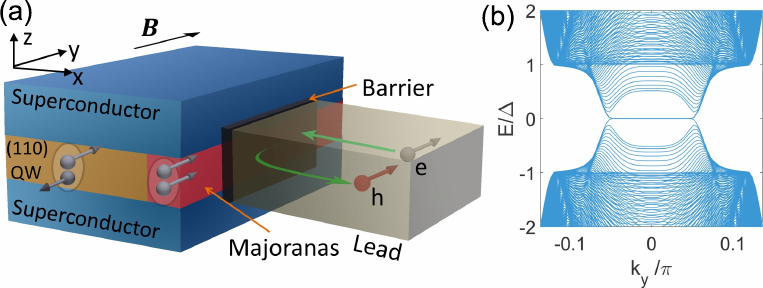}
\caption{(a) The Dresselhaus (110) QW in proximity to s-wave superconductors and subject to a magnetic field $ \bm B $ along $y$ direction. A large number of zero energy Majorana modes associated with MFB are created at the edges parallel to $\bm B$. Electrons can undergo SESAR and inject pairs of electrons with equal spin into the superconductor. Cooper pairs in the bulk are spin-singlet. (b)The band structure of the QW with periodic boundary condition in $y$-direction and open boundary condition in $x$-direction using the tight binding model in Eq.\ref{Htb}. A MFB connects the two nodal points. The parameters in $H_{TB}$ are: $\Delta =1, t=40, \alpha_{D}=30, \mu =-4t, B=1.5$.}
\end{figure}

Specifically, we consider a 2D semiconductor quantum well (QW) grown in the (110) direction with Dresselhaus spin-orbit coupling (SOC) and in proximity to s-wave superconductors \citep{Alicea2010} as depicted in Fig.1. An in-plane Zeeman field can drive the system into a nodal topological phase which supports a large number of zero energy Majorana modes at the edge \citep{JiabinYou,Tanaka110}. This model is considered because many of the nodal topological superconductors studied previously, except Ref.\citep{Chris_MFB}, preserve time-reversal symmetry and cannot induce spin currents. 

In the following sections, we first review the SESAR processes. Secondly, using superconducting (110) Dresselhaus QW as an example, we show that giant spin currents can be injected into nodal topological superconductors using paramagnetic leads. Thirdly, to further explain the origin of the spin current, we show that the pairing correlation at the edge of the Dresselhaus QW is dominantly equal-spin triplet pairing even though superconductivity of the QW is induced by an s-wave superconductor. Finally, we discuss the experimental signatures of the equal-spin triplet Cooper pairs in nodal topological superconductors.

{\bf SESAR and Quantized Spin Conductance} --- A Majorana fermion $\gamma$ is a self-Hermitian particle with the property $ \gamma = \gamma^\dagger$ \citep{Wilczek,Kitaev}. In general, a Majorana fermion can couple to both spin up and spin down electrons. However, due to the self-Hermitian property of the Majorana fermion, the effective coupling Hamiltonian at the N/TS interface can be written as $ i\omega\gamma ( \Psi + \Psi^{\dagger})$ where $\Psi = a \psi_\uparrow + b \psi_\downarrow$ is a linear superposition of spin up and spin down electrons $\psi_{\uparrow/\downarrow}$ in the lead with the normalized coefficients $ |a|^2 +|b|^2 =1 $ and $ \omega>0 $ is the coupling constant. In the basis of $( \psi_{\uparrow}, \psi_{\downarrow}, \psi_{\uparrow}^{\dagger}, \psi_{\downarrow}^{\dagger})$ for the scattering matrix, the Andreev reflection matrix $r^{he}$ and the electron reflection matrix $r^{ee}$ at zero energy can be easily calculated as :
\begin{equation} \label{rhe}
\begin{array}{ccc}
r^{he} &=& \left(
\begin{array}{ cc}
a & b^{*} \\ b & -a^{*}
\end{array} \right)^{*}
\left(
\begin{array}{ cc}
1 & 0 \\ 0 & 0
\end{array} \right)
\left(
\begin{array}{ cc}
a & b^{*} \\ b & -a^{*}
\end{array} \right)^{\dagger}, \\
r^{ee} &=& \left(
\begin{array}{ cc}
a & b^{*} \\ b & -a^{*}
\end{array} \right)
\left(
\begin{array}{ cc}
0 & 0 \\ 0 & 1
\end{array} \right)
\left(
\begin{array}{ cc}
a & b^{*} \\ b & -a^{*}
\end{array} \right)^{\dagger}. 
\end{array}
\end{equation}
It is clear from Eq.\ref{rhe} that $r^{he} (a,b)^{T} = (a^{*}, b^{*})^{T} $ and $r^{he} (b^{*} , -a^{*})^{T} = 0$. It shows that electrons with spinor $ (a, b)$ can undergo resonant Andreev reflections with unity amplitude. Importantly, the reflected hole has spinor $(a^{*}, b^{*})$ due to missing electrons with spinor $(a, b)$ below the Fermi energy. Consequently, a pair of equal spin electrons are injected into the superconductor at each tunnelling event, which results in equal spin Andreev reflections. On the other hand, electrons with orthogonal spinor $ (b^{*}, -a^{*})$ are totally reflected as electrons. This type of Andreev reflection processes is referred to as SESAR \citep{SESAR}.

With the scattering matrices, the charge conductance and the spin conductance in the $\bm{j}$ direction can be calculated as \citep{Datta}:
\begin{eqnarray} 
G_c  = \frac{e^2}{h} \text{tr} \left\lbrace \sigma_0 - r^{ee\dagger} r^{ee}+r^{he\dagger} r^{he} \right\rbrace ,\\  
G_{s, \bm j}  =  \frac{e^2}{h} \text{tr} \left\lbrace - r^{ee\dagger} \bm{j} \cdot \bm\sigma r^{ee}+r^{he\dagger} \bm{j} \cdot \bm\sigma^{*} r^{he}\right\rbrace. 
\end{eqnarray}

Using Eq.\ref{rhe}, we have $G_c = 2\frac{e^2}{h}$ and $G_{s, \bm j}=2\frac{e^2}{h} \bm j \cdot \bm n$, where $\bm n = (a^{*}, b^{*}) \bm \sigma (a, b)^T$ is the spin polarization direction of the electrons undergoing SESAR. In other words, SESAR induces quantized spin conductance in the $\bm n$ direction at zero bias. As long as the coupling $ \omega $ is weakly energy dependent, the current at finite voltage bias is also spin polarized \citep{SESAR}.

%In general, the zero bias Andreev scattering matrix at the interface between a normal lead and a two-channel wire in D class topological superconductor with Majorana end states can be written as []:
%\begin{equation}
%r_{he}= U_{1}^{*}\left(
%\begin{array}{ cc}
%1 & 0 \\ 0 & 0
%\end{array} \right) U_{2}^{\dagger},
%\end{equation}
%where $U_{1}$ and $U_{2}$ are unitary matrices. In this case, an incoming electron with spin $U_{2} (1,0)^{T}$ will be reflected as a hole with spin $U_{1} (1,0)^{T}$. On the other hand, electrons with spin $U_{2} (0,1)^{T}$ will be totally reflected as electrons. Importantly, the zero bias spin conductance can still be quantized due to the selective Andreev reflections. Therefore, Majorana fermion can induce spin polarized current in normal leads.

{\bf MFB in Dresselhaus QW}--- In this section, we consider a zinc-blende (110) quantum well in proximity to s-wave superconductors as depicted in Fig.1a \citep{Alicea2010}. It has been shown that, in the presence of an in-plane magnetic field, such a system can be driven to a nodal topological phase which supports a large number of Majorana modes at the sample edges \citep{JiabinYou, Tanaka110}.

The Hamiltonian of the system, in the Nambu basis of $ (c_{\bm k\uparrow},c_{\bm k\downarrow},c^{\dagger}_{-\bm k\uparrow},c^{\dagger}_{-\bm k\downarrow}) $, can be written as \citep{Alicea2010}: 
\begin{equation} \label{H}
\begin{array} {ll}
H(k_x, k_y)  =  [-2t(\cos k_x +\cos k_y)-\mu]\tau_z \\ 
 +\alpha_D\sin k_x \sigma_z  + B_x\sigma_x\tau_z +B_y\sigma_y +\Delta\sigma_y\tau_y.
\end{array}
\end{equation}
Here $ t $ denotes the hopping amplitude, $ \mu $ is the chemical potential, $ \alpha_D $ is the Dresselhaus SOC strength, $\bm B=(B_x ,B_y ,0) $ is the in-plane magnetic field, and $ \Delta $ denotes the induced superconducting pairing amplitude. $\sigma_i$ and $\tau_i$ are Pauli matrices operating on the spin and particle-hole basis respectively. In this section, the magnetic field is chosen to be in the $y$ direction and $|\bm B|=B$.

Unlike 2D QWs with Rashba SOC $k_x \sigma_y - k_y \sigma_x$, where the electron spins are coupled to both $k_x$ and $k_y$, the electron spins in the Dresselhaus QW couple to $k_x$ only. As a result, for fixed $k_y$, $H(k_x, k_y)= H_{k_y}(k_x)$ is equivalent to the model describing a quantum wire with Rashba SOC strength $\alpha_D$, Zeeman energy $B$, s-wave pairing $\Delta$ and effective chemical potential $\mu'= \mu+2t(1+\cos k_y)$. Therefore, $H_{k_y}(k_x)$ supports Majorana end states when the topological condition $ B^2 > { \mu'^2 + \Delta^2}$ is satisfied \citep{Sato2009,Sau,Alicea2010,ORV}. In one-dimensional wires, it is rather difficult to fine tune the chemical potential to satisfy this topological condition. In the current model, the effective chemical potential is a function of $k_y$ and there is a wide range of $k_y$ such that the topological condition can be satisfied for a given chemical potential. This results in the MFB.

The energy spectrum as a function of $ k_y $ for a 2D quantum well with open boundary conditions in the $x$-direction and periodic boundary conditions in the $y$-direction is depicted in Fig.1b. In calculating the energy spectrum, the following tight-binding model is used:
\begin{eqnarray} \nonumber 
&&H_{TB}=-t\sum_{\bm R,\bm a,s}c^{\dagger}_{\bm R+\bm a,s}c_{\bm R,s}-\frac{\mu}{2}\sum_{\bm R,s}c^{\dagger}_{\bm R,s}c_{\bm R,s}\\ \nonumber
&&+\frac{1}{2}\sum_{\bm R,s,s'}\left[{i\alpha_D}c^{\dagger}_{\bm R+\bm x,s}c_{\bm R,s'}(\sigma_z)_{ss'}
+c^{\dagger}_{\bm R,s}c_{\bm R,s'}(\bm B\cdot\bm\sigma)_{ss'}\right]\\ \label{Htb}
&&+\Delta\sum_{\bm R}c^{\dagger}_{\bm R\uparrow}c^{\dagger}_{\bm R\downarrow}+h.c.,
\end{eqnarray}  
where $ \bm R $ denotes the lattice sites, $ \bm a=\bm x,\bm y $ denotes the primitive vectors along $x$ and $y$ directions, $ s=\uparrow ,\downarrow $ denotes spin.

The bulk gap closes at $k_y$ values where $ B^2 = { \mu'^2 + \Delta^2}$ is satisfied and the nodal points are connected by the MFB as shown in Fig.1b. It is important to note that the MFB realized in this model is protected against short range disorder by a chiral symmetry $C H(k_x,k_y) C^{-1}=-H(k_x,k_y)$, where $C=\sigma_x\tau_y $. This is similar to the case studied in Ref.\citep{Chris_MFB} in which the MFB is created in a $p \pm ip$ superconductor by an in-plane magnetic field.

\begin{figure}
\includegraphics[width=3.2in]{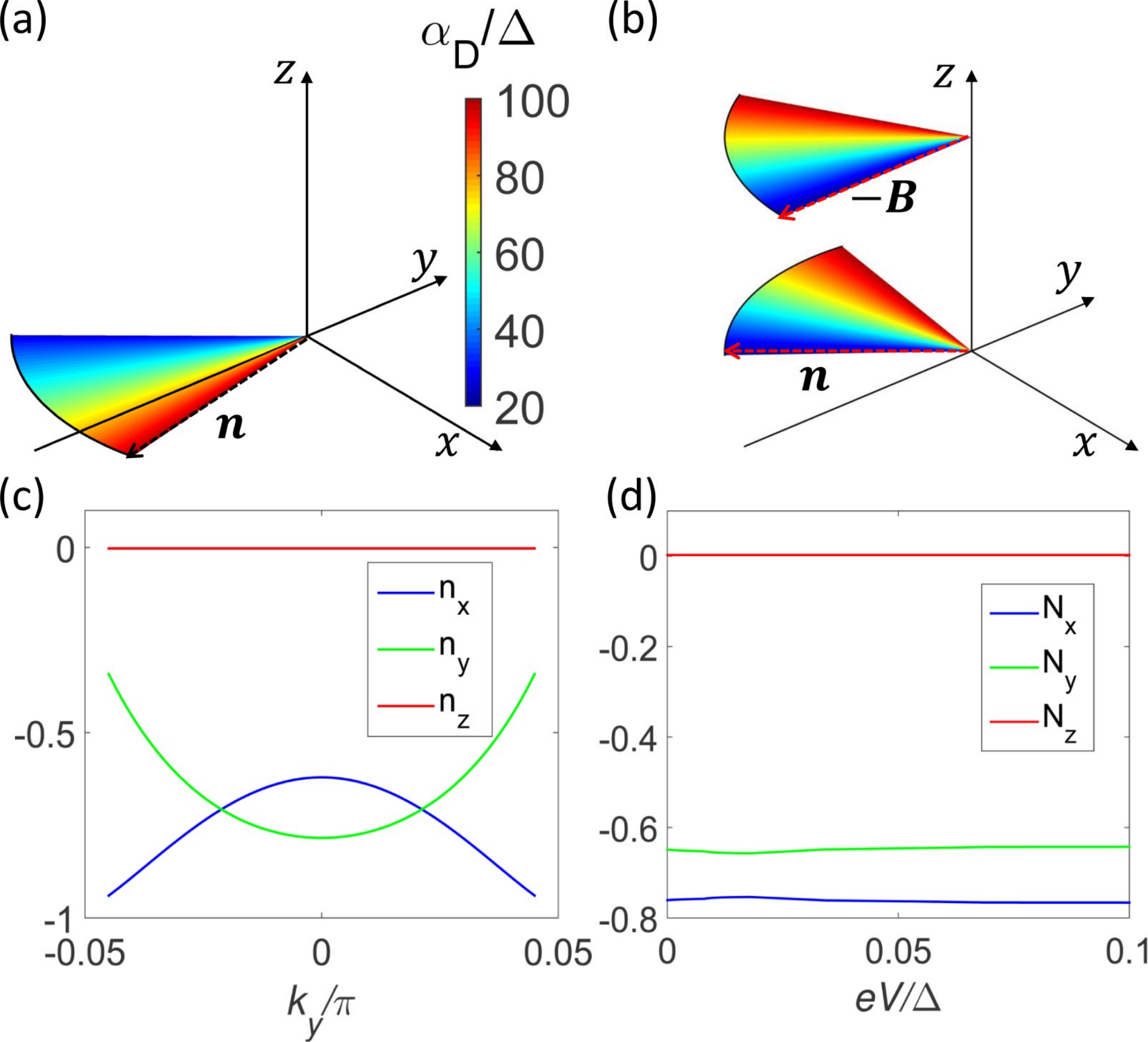}
\caption{(a) The polarization direction $\bm n(k_y =0)$ of the spin current due to SESAR as a function of Dresselhaus SOC strength. The magnetic field $ \bm B $ is along the positive $y$-direction. (b)  $\bm{n}(k_y =0)$ and the in-plane magnetic field directions $-\bm B$. The vector $- \bm B$ with certain colour determines the vector $\bm n$ with the same colour. (c) The three components of $\bm n (k_y=0)$ as a function of $k_y$ in the MFB regime. (d) The components of the spin polarization vector $\bm N$ of the current. The parameters of the QW are the same as Fig.1b. The lead is modelled by square lattices  with hopping $ t_L =2t $. The hopping between the lead and the QW is $t_c =t/10 $.}
\end{figure}

{\bf Giant Spin Currents induced by MFB}--- Since each single Majorana fermion can induce spin currents, we expect that the large number of Majorana modes at the edge of a nodal topological superconductor may induce giant spin currents in leads coupled to the Majorana modes due to resonant Andreev reflections \citep{Vic2009}. However, one has to show that there remains a large spin current after summing up the currents induced by all the Majorana modes.

To proceed, we note that at fixed $k_y$, $H(k_x,k_y)$ in Eq.\ref{H} is in symmetry class BDI and describes a 1D Rashba wire with s-wave pairing \citep{Schnyder_class, ChrisDIII, Sau_BDI}. Assuming periodic in $y$-direction and open boundary conditions in $x$-direction, for a fixed $k_y$ and in the topological regime where the MFB arises, the zero bias Andreev reflection matrix at the interface can be cast into the form \citep{Beenakker2012}:
\begin{equation}
r^{he}_{k_y}= U_{1}^{*}(k_y)\left(
\begin{array}{ cc}
1 & 0 \\ 0 & 0
\end{array} \right) U_{2}^{\dagger}(k_y),
\end{equation}
where $U_{1}$ and $U_{2}$ are $ k_y $-dependent unitary matrices. When the coupling between the lead and the superconductor is weak, $U_{1} = U_{2}$ and the form obtained by effective Hamiltonian approach in Eq.\ref{rhe} is recovered. To obtain $U_1$ and $U_{2}$ for general coupling strengths, the scattering matrix at the N/TS interface at fixed $k_y$ can be calculated as \citep{Lee}:
\begin{eqnarray}\label{S}
\{r_{k_y}^{\alpha \beta}\}_{ij} = - \delta_{ij}\delta_{\alpha \beta} + i\sum_{mn} \{\Gamma_\alpha^{1/2}\}_{im} G^{ \alpha \beta}_{mn}(k_y) \{\Gamma_\beta^{1/2}\}_{nj},
\end{eqnarray}
where $\alpha, \beta \in \{e,h\}$ label the electron or hole, and $i,j,m,n$ label the spin degrees of freedom. $G_{mn}^{\alpha \beta}$ is the retarded Green's function obtained from both $H_{TB}$ in Eq.\ref{Htb} and the Hamiltonian of the lead by assuming periodic boundary conditions in the $y$ direction at fixed $k_y$. $\Gamma_{e/h}$ is the electron/hole part of the broadening function at fixed $k_y$ due to the lead.

In the topological regime with a fixed $k_y$, the channel with spinor $ \Psi_{1}(k_y)=U_{2}(k_y) (0, 1)^{T}$ which undergoes electron reflection can be found by the condition $ r^{he}_{k_y} \Psi_{1}(k_y) = 0$. On the other hand, the channel with spinor $\Psi_{2}(k_y)= U_{2}(k_y) (1,0)^{T}$ undergoes resonant Andreev reflections at zero bias. In the weak coupling limit, the spin polarization of the SESAR process is $\bm n(k_y) = \Psi_{2}^{\dagger}(k_y)\bm\sigma\Psi_{2}(k_y)$.

\begin{figure}
\centering
\includegraphics[height=0.6\textwidth]{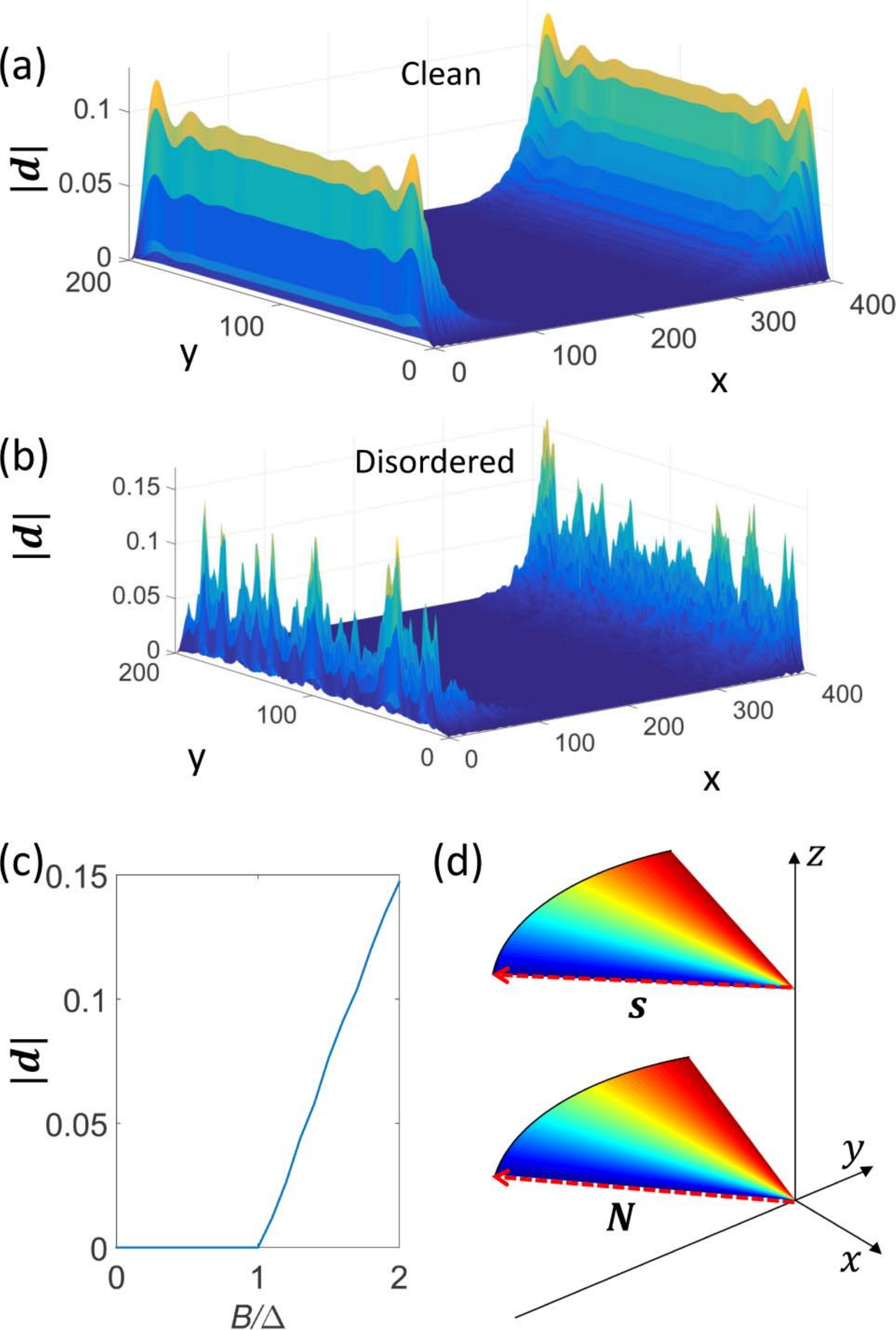}
\caption{(a) and (b) The triplet pairing correlation magnitudes $ |\bm d| $, defined in Eq.10, calculated using $H_{TB}$ in Eq.\ref{Htb} without and with disorder. In (b), the on-site disorder with normal distribution of variance $W=5\Delta$ is added to $H_{TB}$. (c) The spin-triplet pairing correlation magnitude $ |\bm d| $ at site $ (0,100) $ on the edge as a function of $ B/\Delta $. The $\bm d$-vector is non-zero only when the MFB appears.  (d) The Cooper pair spin polarization $\bm s=i(\bm d\times\bm d^{*})/|\bm d|^2$ at site $(0,100)$ on the edge and the spin polarization vector $ \bm N $. Vectors with the same colour indicate the same parameters used in calculating $\bm s$ and $\bm N$.}
\end{figure}

To understand the parameter dependence of the spin polarization direction, $\bm n (k_y=0)$ as a function of Dresselhaus SOC strength $\alpha_D$ is shown in Fig.2a. $\bm n(k_y =0)$ as a function of in-plane magnetic field direction is shown in Fig.2b. The three components of $\bm n$ as functions of $k_y$ are shown in Fig.2c.

At finite voltage bias $ V $, the total spin current in an arbitrary $\bm j$ direction can be worked out by summing up all the $k_y$ components : 
\begin{equation}
I^s_{\bm{j}} = \sum_{k_y \in {(-\pi, \pi]}}\frac{e}{h}\int_{-eV}^{eV}\text{tr}[-r^{ee\dagger}_{k_y}\bm{j}\cdot \bm{\sigma} r^{ee}_{k_y}+r^{he\dagger}_{k_y}\bm{j}\cdot \bm{\sigma}^* r^{he}_{k_y}]dE.
\end{equation}
The spin polarization vector can be defined as $\bm{N} = ( I_{x}^s, I_{y}^s, I_{z}^s )/I_T $ with $I_T = \sqrt{(I_{x}^s)^2 + (I_{y}^s)^2 + (I_{z}^s)^2 }$. The voltage bias dependence of $\bm N$ is depicted in Fig.2d. It is evident that $\bm N$ is almost independent of voltage bias.  As expected, the spin currents induced by the Majorana modes with different $k_y$ do not cancel each other and this results in a spin polarized current.

{\bf Pairing Correlation}--- In the above sections, it is shown using scattering matrices that the normal lead injects pairs of electrons with certain spin polarization into the superconductor to form Copper pairs. On the other hand, the parent superconductor which induces superconductivity in the Dresselhaus QW has s-wave pairing and it is rather surprising that the induced superconductivity on the edge of the Dresselhaus QW is predominantly spin-triplet. 

To further understand the system, we calculate the real space retarded Green's function of the system:
\begin{equation}
G(E) = \left(
\begin{array}{ cc}
G^{ee} & G^{eh} \\ G^{he} & G^{hh}
\end{array} \right) =\frac{1}{E + i0^{+}- H_{TB}}.
\end{equation}
The anomalous part of the retarded Green's function is the Fourier transform of retarded response function:
\begin{equation}
G^{eh}_{s, s'}(E, {\bm R})= -i \int_{0}^{\infty} e^{i(E +i0^{+})t} \langle \{ c_{\bm R,s}(t), c_{\bm R, s'}(0) \} \rangle dt.
\end{equation}
It provides information about the pairing symmetry of the superconductor \cite{Gorkov}. The four spin components of $G^{eh}_{s, s'}$ can be parametrized into the matrix form as:
\begin{equation}
G^{eh}(E, {\bm R}) = (\psi_s + \bm d \cdot \bm \sigma )i\sigma_y.
\end{equation}
Here, $\psi_s =$tr$[-i G^{eh}\sigma_y]/2$ gives the spin-singlet pairing correlation $\langle \psi^{\dag}_{\uparrow}\psi^{\dag}_{\downarrow}- \psi^{\dag}_{\downarrow}\psi^{\dag}_{\uparrow}\rangle $ and the $\bm d$-vector characterizes the spin-triplet pairing correlation. For example, $d_{x} = $tr$[-i\sigma_x G^{eh}\sigma_y]/2$ is the expectation value of $ \psi^{\dag}_{\uparrow}\psi^{\dag}_{\uparrow}- \psi^{\dag}_{\downarrow}\psi^{\dag}_{\downarrow}$. The retarded Green's function is calculated based on the tight-binding model in Eq.\ref{Htb} by recursive approach \citep{Lee}. The resulting position dependence of the triplet pairing correlation $ |\bm{d}|$ at zero energy is displayed in Fig.3a. The spin-singlet pairing correlation at zero energy is negligible in the whole sample. It is important to note that the triplet pairing correlation is strongest near the edges where the Majorana fermions reside. 

Moreover, since the Majorana modes are protected from disorder by a chiral symmetry \citep{Chris_MFB}, the spin-triplet pairing correlations survive even if onsite disorder is introduced into the sample as shown in Fig.3b. The spin-triplet pairing correlation $|\bm d|$ of a chosen site as a function of Zeeman energy $B$ is shown in Fig.3c. The spin-triplet pairing correlation is finite only when the superconductor enters the nodal topological phase with $B > \Delta$.

\begin{figure}
\includegraphics[width=3.2in]{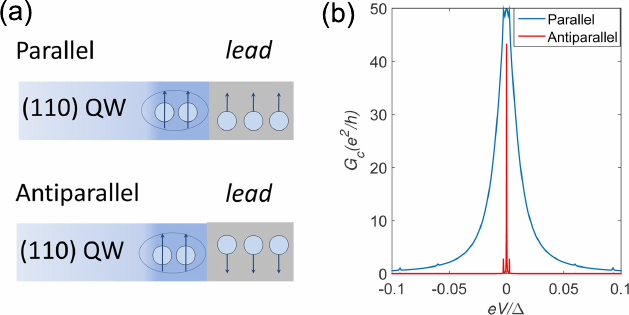}
\caption{(a) The tunnelling between a half metal lead and the QW. The spin polarization of the lead $\bm h$ are approximately parallel or antiparallel to the Cooper pair spin polarization $\bm s$. (b) The corresponding charge conductance $ G_c $ in the parallel ($\bm h \parallel \bm s$) and antiparallel ($\bm h \parallel -\bm s$). Here the parameters used are the same as Fig.2d, except that a Zeeman field with magnitude $ |\bm h| = 2\Delta $ is added to the lead to polarize the electron spins in the lead. The hopping between the lead and the QW is $t_c =t/2 $.}
\end{figure}

Interestingly, from the $\bm d$-vector, the spin polarization direction of the Cooper pair is found to be $ \bm s=i(\bm d\times\bm d^{*})/|\bm d|^2 $\citep{Leggett}. It is shown in Fig.3d that the spin polarization direction of the Cooper pairs matches the spin polarization of the tunnelling current found in Fig.2d. 

Since the Cooper pairs are spin-polarized in the $\bm s$ direction, a half-metal lead (such as CrO$_{2}$\citep{Gupta}) with spin polarization $\bm h$ parallel to the Cooper pair spin polarization direction $\bm s$ can freely inject Cooper pairs into the superconductor. The $\bm h $ dependence of the tunnelling current from a half-metal lead to the nodal topological superconductor is shown in Fig.4. In Fig.4b, it is shown that the zero bias conductance peak (ZBCP) is very wide when $\bm h$ is approximately parallel to the spin polarization direction of the Cooper pair. Due to the large number $m$ of Majorana modes at the edge, the ZBCP can be as large as $2m\frac{e^2}{h}$ due to Majorana induced resonant Andreev reflections \citep{Vic2009}. Practically, the ZBCP is limited by the number of conducting channels in the lead. On the other hand, the width of the ZBCP is greatly suppressed when $\bm h$ is approximately antiparallel to the spin polarization direction of the Cooper pairs. This feature can be used to detect the Majorana fermions in nodal superconductors.

{\bf {Conclusion}}--- In this work, we show that giant spin current can be injected into nodal topological superconductors using paramagnetic leads due to SESAR. SESAR is related to the spin-triplet correlations at the edge of the topological superconductors which can be detected by tunnelling experiments.

The authors thank the support of HKRGC and Croucher Foundation through HKUST3/CRF/13G, 602813, 605512, 16303014 and Croucher Innovation Grant.

\end{document}